\documentclass[aps,prl,10pt,twocolumn,superscriptaddress]{revtex4-2}
\newcommand{\beginsupplement}{
  \clearpage
  \onecolumn
  \setcounter{section}{0}
  \setcounter{equation}{0}
  \setcounter{figure}{0}
  \setcounter{table}{0}
  \renewcommand{\thesection}{S\Roman{section}}
  \renewcommand{\theequation}{S\arabic{equation}}
  \renewcommand{\thefigure}{S\arabic{figure}}
  \renewcommand{\thetable}{S\arabic{table}}
}
\usepackage{amsmath,amssymb}
\usepackage{graphicx}
\usepackage[usenames]{xcolor}
\usepackage{amssymb, amsmath}
\definecolor{columbiablue}{RGB}{0,114,206}
\usepackage[colorlinks,allcolors=columbiablue]{hyperref}
\usepackage{amsmath}
\usepackage{bm}
\usepackage{float}
\usepackage[utf8]{inputenc}
\usepackage{braket}
\usepackage[dvipsnames]{xcolor}
\graphicspath{{Figures/}}
\usepackage[normalem]{ulem}

\def\bi{{\bf i}}
\def\bj{{\bf j}}
\def\bk{{\bf k}}

\begin{document}

\title{Hole and spin dynamics in an anti-ferromagnet close to half filling}
\author{Magnus Callsen}
\affiliation{Department of Physics and Astronomy, Aarhus University, Ny Munkegade 120, DK-8000 Aarhus C, Denmark}
\author{Jens H.\ Nyhegn}
\affiliation{Kavli Institute for Theoretical Physics, University of California, Santa Barbara, California 93106-4030, USA}
\affiliation{Niels Bohr Institute, University of Copenhagen, Jagtvej 128, DK-2200 Copenhagen, Denmark}
\author{Kristian Knakkergaard Nielsen}
\affiliation{Niels Bohr Institute, University of Copenhagen, Jagtvej 128, DK-2200 Copenhagen, Denmark}
\author{Georg M.\ Bruun}
\affiliation{Department of Physics and Astronomy, Aarhus University, Ny Munkegade 120, DK-8000 Aarhus C, Denmark}
\date{\today}

\begin{abstract} 
The interplay between charge  and spin dynamics is at the heart of  strongly correlated materials. Inspired by recent quantum simulation experiments, we develop a conserving diagrammatic method to describe the Fermi-Hubbard model for strong repulsion and small hole doping away from the half-filled anti-ferromagnetic ground state. We show that doping leads to four hole pockets in the Brillouin zone formed by magnetic polarons, which become increasingly damped with hole concentration. Likewise, the magnon spectrum of the anti-ferromagnet softens and dampens with doping due to hole-induced magnetic frustration. This gives rise to  a suppression of the anti-ferromagnetic correlations in agreement with recent experiments. We then calculate the response of the system to a lattice modulation and recover the qualitative difference between in-phase and out-of-phase modulations seen in  experiments, which was interpreted as signs of pseudogap physics. Our results indicate that the complex competition between spin and charge degrees of freedom and the emergence of the pseudogap phase may be usefully analyzed for small dopings, where systematic theories can be developed.  
\end{abstract}

\maketitle
Analyzing charge motion in a background of strongly correlated spins represents a key challenge for understanding two-dimensional (2D) materials, where 
it gives rise to a 
wealth of phenomena including charge and density order as well as  high-temperature ($T_c$) superconductivity. While some features depend on the specific material,
many are robust and  exist across  different families of compounds~\cite{Keimer:2015aa}. 
The single band Fermi-Hubbard model  offers a minimal model for describing such generic effects in strongly correlated materials, and its phase diagram continues to be intensely 
investigated numerically~\cite{LeBlanc2015,Arovas2022}. Quantum simulation experiments with  atoms in optical lattices provide a powerful alternative 
way to explore the Fermi-Hubbard model under pristine conditions~\cite{bakr2025microscopyultracoldfermionsoptical,Koepsell2019,Koepsell2021,Ji2021,Hartke2023,Xu:2023aa,Prichard2024,Lebrat2024}, and 
two such experiments have recently reported intriguing results regarding the effects of  hole doping. In the first, spin and charge correlations were observed  with quantum gas microscopy~\cite{Chalopin2026}. In the second, 
lattice modulation spectroscopy combined with a new powerful cooling technique  was used to measure the particle-hole excitation spectrum in the first Brillouin zone (BZ)~\cite{kendrick2025pseudogapfermihubbardquantumsimulator}. Both experiments were interpreted in terms of pseudogap physics where the density-of-states is suppressed at the Fermi level in certain regions of the BZ. The pseudogap is observed in many 2D materials including the cuprates both in transport~\cite{Chan2025}  and in angle-resolved photoemission spectroscopy where it gives rise to the famous Fermi arcs~\cite{Sobota2021}, but its physical origin remains highly debated and it is believed to hold the key for understanding high-$T_c$ superconductivity~\cite{Keimer:2015aa,Bonetti2025}.

Here,  we theoretically investigate the single band nearest neighbour Fermi-Hubbard model for small hole dopings away from half-filling (one fermion per site) where the ground state exhibits long range anti-ferromagnetic (AFM) order.  Using a conserving diagrammatic approach based on a self-consistent Born approximation combined with a random phase approximation 
including anomalous propagators, we show that  hole doping  gives rise to the formation of magnetic polarons residing in 
four elliptical hole pockets in the BZ. The magnon (spin wave) spectrum softens with hole doping and both the magnons and 
the magnetic polarons become increasingly damped. This is shown to lead to a decrease 
in the AFM correlations,  and we finally analyze the response of the system to lattice modulations. Our results agree  with recent quantum simulation experiments~\cite{Chalopin2026,kendrick2025pseudogapfermihubbardquantumsimulator}, suggesting that the competition between charge motion and spin correlations and the onset of pseudogap physics can be usefully analyzed from the small doping side. 

\paragraph{Model.--}
Consider spin $\sigma=\uparrow,\downarrow$ fermions in a square lattice as described by the Fermi-Hubbard model with nearest neighbour hopping $t$ and on-site 
interaction  $U$. For strong repulsion $U/t\gg 1$ and close to half-filling, 
the low-energy physics is well-described by the $t$-$J$ model~\cite{Auerbach_book} 
\begin{align}
\hat H = -t\sum_{\langle \bi,\bj\rangle,\sigma} \!\!\Big(\tilde{c}_{\bi\sigma}^\dagger \tilde{c}_{\bj\sigma} \!+\! \text{h.c.}\Big) \!+\! J\sum_{\langle \bi,\bj\rangle}\!\!\Big(\hat{\mathbf{S}}_\bi \cdot \hat{\mathbf{S}}_\bj \!-\! \frac{\hat n_\bi\hat n_\bj}{4}\Big), 
\label{t-j}
\end{align}
where $\langle \bi,\bj\rangle$ denotes nearest neighbour lattice sites,  $\tilde {c}^\dagger_{i\sigma}=\hat c^\dagger_{\bi\sigma}(1-\hat n_\bi)$ with $\hat c^\dagger_{\bi\sigma}$
creating a fermion with spin $\sigma$ at site $\bi$, and $\hat n_\bi=\hat c^\dagger_{\bi\uparrow}\hat c_{\bi\uparrow}+\hat c^\dagger_{\bi\downarrow}\hat c_{\bi\downarrow}$ the number of fermions at site $\bi$. The spin operators are $\hat{ {\bf S} }_{\bi} = \frac{1}{2}\sum_{\sigma,\sigma'} \hat{ c }^\dagger_{{\bi},\sigma}\boldsymbol{\sigma}_{\sigma\sigma'}\hat{ c }_{{ \bi},\sigma'}$ with the Pauli matrices $\boldsymbol{\sigma}=(\sigma_x,\sigma_y,\sigma_z)$, and  $J=4t^{2}/U$  is the anti-ferromagnetic super-exchange   coupling. 

Close to half-filling, the ground state is an AFM with spins on sublattices $\bi\in A$ and $\bj\in B$ predominantly pointing parallel/anti-parallel to the $z$-direction, respectively. 
Using a Holstein-Primakoff transformation generalized to take  holes into account, we write  
$\hat S^z_\bi=1/2-\hat h_\bi^\dagger\hat h_\bi/2-\hat s_\bi^\dagger\hat s_\bi$, 
$\hat S_\bi^+=(1-\hat h_\bi^\dagger\hat h_\bi-\hat s_\bi^\dagger \hat s_\bi)\hat s_\bi$, 
$\tilde c_{\bi\downarrow}=\hat h_\bi^\dagger\hat s_\bi$, and $\tilde c_{\bi\uparrow}=\hat h_\bi^\dagger(1-\hat s_\bi^\dagger\hat s_\bi)$ for sites on  sublattice 
$\bi\in A$. Here, $\hat h_\bi/\hat s_\bi$ is a fermionic/bosonic operator removing a spinless hole/charge-less spin excitation  at site $\bi$, and 
 we have used a Newton 
expansion~\cite{Konig2021} to simplify the usual slave fermion representation~\cite{Kane1989}. The transformation on  sublattice $\bj\in B$  is obtained from the above by swapping spin $\uparrow$ and $\downarrow$. Keeping terms up to third order in $\hat h_\bi$ and $\hat s_\bi$, Eq.~\eqref{t-j} becomes~\cite{Kane1989,Nielsen2021}
\begin{align}
    \hat H \simeq\sum_{\mathbf{k}}\omega_\mathbf{k}\hat \sigma_{\mathbf{k}}^\dagger \hat\sigma_{\mathbf{k}}+ \sum_{\mathbf{k}, \mathbf{q}}
g(\mathbf{q}, \mathbf{k}) (\hat h_{\mathbf{k}+\mathbf{q}}^\dagger\hat h_{\mathbf{q}}\hat \sigma_{-\mathbf{k}}^\dagger + \text{h.c.}),
\label{tjspinonmagnon}
\end{align}
where we omit the constant ground state energy term. Here, $\hat\sigma_{\mathbf k}^\dagger $ creates a bosonic magnon with crystal momentum $\mathbf k$ 
and energy $\omega_{\mathbf k}=2J \sqrt{1-\gamma_{\mathbf k}^2}$ where 
$\gamma_{\mathbf k}=(\cos k_x+\cos k_y)/2$ within linear spin-wave theory (LSWT)~\footnote{We use units where the reduced Planck constant, the lattice spacing, and 
the Boltzmann constant are set to unity, $\hbar, a, k_B = 1$}. The vertex $g(\mathbf{q}, \mathbf{k})=4tN^{-1/2} (u_{\mathbf k}\gamma_{\mathbf k+\mathbf q}-v_{\mathbf k}\gamma_{\mathbf q})$ gives the amplitude for the emission/absorption of magnons by the holes, which clearly shows the competition between charge motion and magnetic order. 
Here, $u_{\mathbf k}=([(1-\gamma_{\mathbf k}^2)^{-1/2}+1]/2)^{1/2}$, 
$v_{\mathbf k}=\text{sgn}(\gamma_{\bk})([(1-\gamma_{\mathbf k}^2)^{-1/2} - 1]/2)^{1/2}$, and  $N$ is the number of lattice sites. Finally, $\hat h_{\mathbf k}=N^{-1/2}\sum_{\bi}e^{-i{\mathbf k}\cdot \bi}\hat h_\bi$ removes a hole with momentum ${\mathbf k}$.

\paragraph{Diagrammatic theory.--}
The Hamiltonian in Eq.~\eqref{tjspinonmagnon} cannot be solved exactly and we must, therefore, resort to approximations to analyze the charge and spin dynamics. To do this, we employ a diagrammatic approach based on the finite temperature hole Green's function $G({\mathbf k},\tau)=-\braket{ T_\tau\{\hat h_{\mathbf k}(\tau),\hat h_{\mathbf k}^\dagger(0)\}}$. Here, $\hat a(\tau) = e^{+\hat H\tau}\hat a e^{-\hat H\tau}$ with the imaginary time $\tau\in [-1/T,1/T[$ for temperature $T$, $T_\tau$ is time ordering, and $\braket{\cdots}$ denotes the 
thermal expectation value at a given hole concentration. We calculate the hole Green's function using a self-consistent Born approximation (SCBA), which includes all non-crossing diagrams for the  self-energy to infinite order. The SCBA was first introduced to describe a single hole in a half-filled AFM~\cite{Kane1989}, where it turns out to agree remarkably well with Monte-Carlo calculations~\cite{Diamantis2021}, and it has recently been shown to accurately describe non-equilibrium quantum simulation
experiments as well~\cite{Nielsen2022_2}. Here, we generalize the SCBA to the case of a non-zero hole concentration. To describe the effects of the holes on the magnons, we define the normal and anomalous  magnon Green's functions $D_{11}({\mathbf k},\tau)=-\langle T_\tau\{\hat \sigma_{\mathbf k}(\tau),\hat \sigma_{\mathbf k}^\dagger(0)\}\rangle$, $D_{12}({\mathbf k},\tau)=-\langle T_\tau\{\hat \sigma_{\mathbf k}(\tau),\hat \sigma_{-\mathbf k}(0)\}\rangle$ and
$D_{21}({\mathbf k},\tau)=-\langle T_\tau\{\hat \sigma_{-\mathbf k}^\dagger(\tau),\hat \sigma^\dagger_{\mathbf k}(0)\}\rangle$. The anomalous propagators arise because magnon pairs are created by the holes.
We employ a self-consistent random-phase-approximation (RPA) for the magnon self-energy. The resulting Dyson equations are shown diagrammatically in Fig.~\ref{fig:Feyn}(a), and 
the mathematical expressions are given in Supplemental Material~\cite{SM}, where it is also shown that our approach is conserving \cite{Baym1961,Baym1962,Stefanucci_vanLeeuwen_2025}. The electron filling fraction is $\langle\hat n_i\rangle=1-\langle\hat h_i^\dagger\hat h_i\rangle=1-\delta$, where the hole doping 
$\delta=N^{-1}\sum_{\mathbf k}\langle\hat h_{\mathbf k}^\dagger\hat h_{\mathbf k}\rangle=N^{-1}\sum_{\mathbf k}\int_{-\infty}^\infty\!d\omega f(\omega)A_h({\mathbf k},\omega)$ is controlled by 
the hole chemical potential $\mu$. Here, $A_h({\mathbf k},\omega)=-2\text{Im}G({\mathbf k},\omega)$ is the hole spectral function obtained by the  analytical continuation $i\omega_n\rightarrow \omega+i0_+$, and  $f(\omega)=1/[\exp(\omega/T)+1]$ is the Fermi-Dirac function.

The  $t-J$ model 
has been studied for small hole dopings using similar approaches based on the SCBA and the RPA, combined with additional 
approximations for the Green's functions~\cite{Igarashi1992,Khaliullin1993,Sherman1993} or with the omission of anomalous magnon propagators~\cite{Satyaki2011}. Compared to this, we present a fully self-consistent treatment including anomalous propagators, 
and our focus on new quantum simulation experiments that probe the pristine single-band Fermi–Hubbard model  in 
ways beyond the reach of solid-state experiments, is also quite different from these earlier works. 
In the following, we assume zero temperature $T=0$ and take $U/t=7$ giving $J/t=4/7$, which  is equal or close to what is  realised in the  recent quantum simulation experiments~\cite{kendrick2025pseudogapfermihubbardquantumsimulator,Chalopin2026} and also similar to what is found in the cuprates~\cite{Keimer:2015aa}. We solve the Dyson equations iteratively on a $20\times 20$ lattice until convergence is obtained adding a small imaginary part $\eta=0.04t$ to the frequencies. Both the hole and magnon spectral functions fulfill $\int\!d\omega A(\omega)/2\pi=1$ within a few percent for the 
small dopings explored here, illustrating the conserving nature of our theory and accuracy of the numerics.

 \begin{figure}[t!]
    \centering
    \includegraphics[width=\columnwidth]{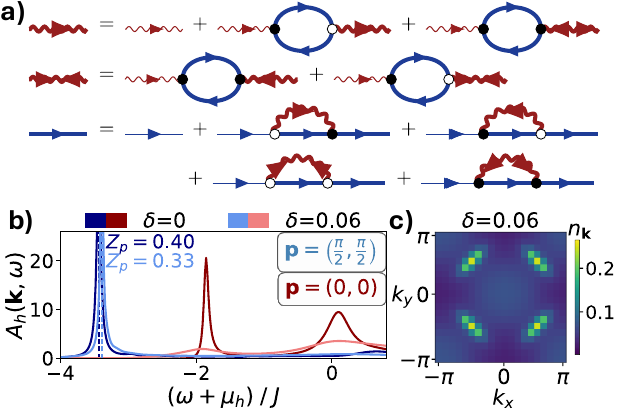}
    \caption{(a) Feynman diagrams for the conserving approximation: thick (thin) straight and wavy lines denote full (bare) hole and magnon propagators, respectively. Oppositely directed arrows indicate anomalous magnon propagators. (b) Hole spectral function at momenta ${\mathbf p}=(0,0)$ and $(\pi/2,\pi/2)$ for dopings $\delta=0$ and $0.06$. (c) Hole occupation in the BZ for $\delta=0.06$.}
    \label{fig:Feyn}
\end{figure}   
 

\paragraph{Holes and magnons.--}
In Fig.~\ref{fig:Feyn}(b), we show the hole spectral function at half-filling ($\delta=0$) and hole doping $\delta=0.06$ at  crystal momenta ${\mathbf p}=(0,0)$ and $(\pi/2,\pi/2)$. For  zero doping, there is a prominent quasiparticle peak for both momenta corresponding to magnetic polarons with energies $\omega\simeq-1.9J$ and $\omega\simeq-3.4J$. The ${\mathbf p} =(\pi/2,\pi/2)$ polaron remains well-defined also for  $\delta=0.06$ with a slightly higher energy and smaller residue, whereas the ${\mathbf p}=(0,0)$ polaron becomes overdamped. The reason is that a finite concentration of holes causes magnetic frustration corresponding to a finite magnon population on which the holes can scatter.
This most strongly affects the polaron for ${\mathbf p}=(0,0)$ where the energy is maximal, whereas  magnetic polaron is more robust 
for ${\mathbf p}=(\pm \pi/2,\pm \pi/2)$ where we find its energy to be  minimal   consistent with earlier findings for zero doping~\cite{Kane1989,Diamantis2021,Nielsen2021,Massignan2025}.

Figure~\ref{fig:Feyn}(c) plots the hole population $\langle\hat h_{\mathbf p}^\dagger\hat h_{\mathbf p}\rangle$
in the BZ for the doping $\delta=0.06$. This clearly shows the presence of four elliptic hole pockets centered around the momenta $(\pm \pi/2,\pm \pi/2)$~\cite{Satyaki2011,KYUNG1998475}, 
which can  be understood as arising from Fermi seas of magnetic polarons.

\begin{figure}[t!]
    \centering
    \includegraphics[width=\columnwidth]{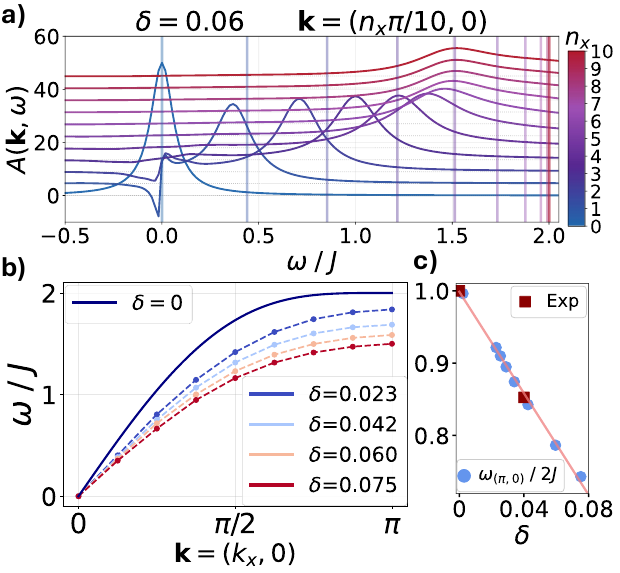}
    \caption{(a) Magnon spectral function for hole doping $\delta=0.06$ and different momenta along the $k_x$ direction in the BZ. 
    Vertical lines give the magnon energies at $\delta=0$. (b) Magnon dispersion for ${\bf k}=(0,0)$ to ${\bf k}=(\pi,0)$ in steps of $\pi/10$ and different hole dopings. (c) The magnon energy for ${\bf k}=(\pi,0)$ as a function of $\delta$
    in units of its value for $\delta=0$. The red squares show how the  frequency position of the experimentally measured peak response to an out-of-phase lattice modulation decreases with 
    doping relative to its value     at $\delta=0$~\cite{kendrick2025pseudogapfermihubbardquantumsimulator}. 
    }
    \label{fig:Magnonspectral}
\end{figure}

Figure \ref{fig:Magnonspectral}(a) shows the magnon spectral function $A_{11}({\mathbf k},\omega)=-2\text{Im}D_{11}({\mathbf k},\omega)$
for different momenta along the $k_x$-direction and hole doping $\delta=0.06$. This shows that the presence of holes gives rise to a softening and broadening of the 
magnon dispersion, whose width vanishes for zero doping where the magnons have an infinite life-time within LSWT. 
 The significant spectral weight around $\omega=0$ for small momenta has been suggested to indicate an instability to a stripe or spiral order~\cite{Satyaki2011}.  We 
plot in Fig.~\ref{fig:Magnonspectral}(b) the magnon dispersion along the $k_x$-direction  as obtained from the peak positions of 
 $A_{11}(\omega)$ for different dopings, and 
 in Fig.~\ref{fig:Magnonspectral}(c) the maximum magnon energy at ${\bf k}=(\pi,0)$ is shown. 
 Both plots further illustrate  the softening of the magnon spectrum. 

\paragraph{Spin correlations.--}
We  now turn our attention to the  spin correlations, 
which were recently probed using 
quantum gas microscopy~\cite{Chalopin2026}. Due to the SU(2) spin symmetry in the experiment, the magnetic order at $T=0$ will point in an arbitrary direction from shot to shot. 
To take  this into account, we therefore calculate the averaged spin-spin correlator 
\begin{align}
{\mathcal C}_2({\mathbf d})= \frac{4}{3}\big(\langle\hat{ \mathbf{S}}_{\mathbf{i}+\mathbf{d}} \cdot \hat{ \mathbf{S}}_\mathbf{i} \rangle - \langle \hat{ \mathbf{S}}_{\mathbf{i}+\mathbf{d}}\rangle\cdot\langle\hat{ \mathbf{S}}_\mathbf{i}\rangle\big). 
\label{C2}
\end{align}
A consistent calculation of ${\mathcal C}_2({\mathbf d})$ for a non-zero doping is very challenging. It is however known that LSWT agrees very well with Monte-Carlo results for $\langle\hat S_{\bf i}^z\rangle$~\cite{Sandvik1997,Castilla1991}, and higher order corrections are also found to be rather small for the equal time spin correlations for $\delta=0$, especially for high momenta corresponding to small distances $d$~\cite{Igarashi1991}. We, therefore, calculate $\mathcal{C}_2(d)$ keeping only the dominant lowest order terms in the hole and magnon populations as detailed in Supplemental Material~\cite{SM}.

\begin{figure}[t!]
    \centering
    \includegraphics[width=\columnwidth]{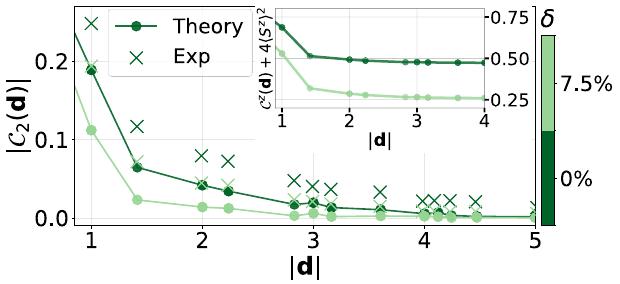}
    \caption{The  correlation function ${\mathcal C}_2({\mathbf d})$ for two dopings; 
    crosses denote experimental data from 
    Ref.~\cite{Chalopin2026}. For clarity,  ${\mathcal C}_{2}({\mathbf 0})=1-\delta$ is not shown. The inset 
    shows  $4\langle\hat S_{\mathbf{i}+\mathbf{d}}^z\hat S_\mathbf{i}^z\rangle$. 
     }
    \label{fig:Spincorrelations}
\end{figure} 
Figure \ref{fig:Spincorrelations} plots ${\mathcal C}_2({\mathbf d})$ for two  different dopings together with the experimental results~\cite{Chalopin2026}. 
We see that the correlations decrease from the value $1-\delta$ at $d=0$  over 
a few lattice distances. The theory predicts this decrease  to be non-monotonic with a small maximum at $d=3$, which we attribute to the 
fact that ${\mathcal C}_2({\bf d})$ is $C_4$ and not circular symmetric. There is reasonable agreement between our theory and the experiment, and 
in both cases the correlations decrease with doping.  The 
 experimental data lie consistently above our theory, which is likely because the experiment was conducted at the  temperature $T/t=0.27$ where there is no long range AFM order. 
In the inset, we plot $\langle\hat{S}_{\mathbf{i}+\mathbf{d}}^z\hat{S}_\mathbf{i}^z\rangle$, which decreases with $\delta$ reflecting the decrease in magnetic order. As detailed in Supplemental Material\cite{SM}, this gives give rise to a decrease in the magnetic structure factor in agreement with the experiment~\cite{Chalopin2026}. 

\paragraph{Lattice modulation spectroscopy.--}
The suppression of the density-of-states at the Fermi level along the nodal directions of the BZ is a hallmark of the pseudogap phase~\cite{Keimer:2015aa}. Lattice modulation spectroscopy with $^6$Li atoms in an optical lattice was recently used to explore this in the Fermi-Hubbard model close to half filling~\cite{kendrick2025pseudogapfermihubbardquantumsimulator}. These experiments probe the imaginary part of the retarded correlation function  $\chi_{\pm}(\omega)$, where $\omega$ is the angular frequency  of the modulation and $\chi_{\pm}(\omega)$ 
the Fourier transform of $\chi_{\pm}(\tau)=-i\theta(\tau)\langle[\hat T_{\pm}(\tau),\hat T_{\pm}(0)]\rangle$ 
(here $\tau$ denotes real time). The modulation operator is  
\begin{align}
\hat T_{\pm}=\sum_{{\bf l}\sigma}(\hat c_{{\bf l}\sigma}^\dagger \hat c_{{\bf l}+{\bf e}_x\sigma} \pm \hat c_{{\bf l}\sigma}^\dagger \hat c_{{\bf l}+{\bf e}_y\sigma})+\text{h.c.}
\label{Shaking1}
\end{align}
with ${\bf l}+{\bf e}_x$  the nearest neighbour  site in the $x$-direction (likewise for ${\bf l}+{\bf e}_y$).  The $\pm$ sign corresponds to 
 the $x$ and $y$ lattice 
depths modulated in- or out-of-phase~\cite{kendrick2025pseudogapfermihubbardquantumsimulator}.

  In the slave-fermion representation, Eq.~\eqref{Shaking1} reads 
\begin{align}
\hat T_+^\text{eff} &= \sum_{\mathbf{k}, \mathbf{q}}g_+(\mathbf{q}, \mathbf{k}) (\hat h_{\mathbf{k}+\mathbf{q}}^\dagger\hat h_{\mathbf{q}}\hat \sigma_{-
\mathbf{k}}^\dagger \!+\! \text{h.c.}) \!+\! \frac2t\sum_{\mathbf{k}}\omega_\mathbf{k}\hat \sigma_{\mathbf{k}}^\dagger \hat\sigma_{\mathbf{k}}\nonumber\\
\hat T_-^\text{eff} &= \sum_{\mathbf{k}, \mathbf{q}}g_-(\mathbf{q}, \mathbf{k}) (\hat h_{\mathbf{k}+\mathbf{q}}^\dagger\hat h_{\mathbf{q}}\hat \sigma_{-\mathbf{k}}^\dagger \!+\! \text{h.c.})\nonumber\\
&+\frac{2J}{t}\sum_{\mathbf k}\frac{\tilde\gamma_{\mathbf k}}{\sqrt{1-\gamma_{\mathbf k}^2}}(\hat \sigma_{\mathbf k}\hat\sigma_{-\mathbf k} \!+\! \hat \sigma^\dagger_{-\mathbf k}\hat\sigma^\dagger_{\mathbf k} \!-\! 2\gamma_{\mathbf k}\hat \sigma^\dagger_{\mathbf k}\hat\sigma_{\mathbf k}),\label{Shaking}
\end{align}
where $g_+(\mathbf{q}, \mathbf{k}) = g(\mathbf{q}, \mathbf{k})/t$, $g_-(\mathbf{q}, \mathbf{k}) = 4 N^{-1/2}(u_{\mathbf k}\tilde\gamma_{\mathbf k+\mathbf q}-v_{\mathbf k}\tilde\gamma_{\mathbf q})$, and $\tilde \gamma_{\mathbf k}=(\cos k_x - \cos k_y)/2$. 
We have defined $\hat T_\pm^\text{eff}=\hat P_Le^{\hat S}\hat T_\pm e^{-\hat S}\hat P_L$ where 
 $e^{\hat S}$ is the unitary Schrieffer-Wolf transformation, which together with the projection   $\hat P_L$   onto states with maximally one 
particle per site is used to map the Hubbard model onto the $t-J$ model, see Supplemental Material \cite{SM}.

\begin{figure}[t!]
    \centering
    \includegraphics[width=\columnwidth]{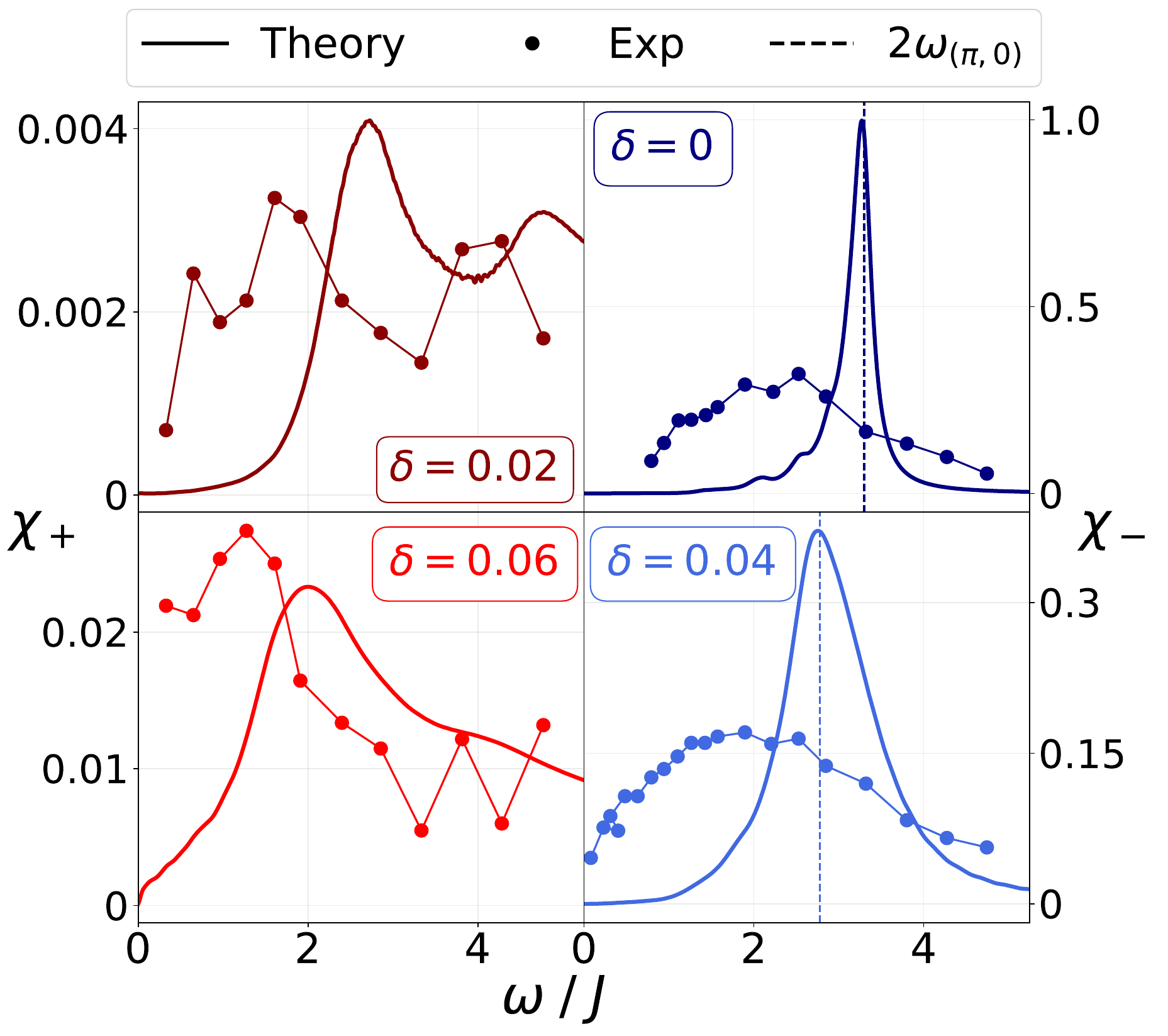}
   \caption{ 
   The left/right  column show the response $\chi_{\pm}(\omega)$ to an in-phase/out-of-phase lattice modulation for different dopings normalised by the maximal value for $\delta=0$. Vertical lines indicate the maximal energy $2\omega_{(\pi,0)}$ required to create two magnons. Data points are 
   experimental results from Ref.~\cite{kendrick2025pseudogapfermihubbardquantumsimulator} normalised to have the same area as the theoretical curves. 
 }
    \label{fig:Modulation}
\end{figure}

A conserving calculation of  $\chi_{\pm}(\omega)$ is challenging, and  we therefore use perturbation theory instead keeping the lowest order contributions
in the hole and magnon populations as explained 
in Supplemental Material \cite{SM}. The left and right columns of Fig.~\ref{fig:Modulation} plot the response to an in-phase/out-of-phase
 lattice modulation respectively for different dopings. 
Consider first $\delta=0$ where the out-of-phase response comes only from the creation of magnon pairs due to  
the $\hat \sigma^\dagger_{-\mathbf k}\hat\sigma^\dagger_{\mathbf k}$ term in 
Eq.~\eqref{Shaking}. The maximum of this response is at  $\omega=4J$  corresponding to the creation of two magnons with 
maximal energy $2J$. As discussed in  connection with solid-state Raman scattering experiments~\cite{Devereaux2007,Priv_Comm_Eugene_Demler}, since the magnons are created in pairs, interactions between them reduce their energy to $3.3J$. This effect is not taken into account by LSWT, and we, therefore, include it approximately by rescaling the spin coupling as $4J\rightarrow3.3J$, when calculating the response. Note that the response to an in-phase lattice modulation  vanishes for  $\delta=0$ (not shown), since  this simply corresponds to a global modulation of  the spin coupling $J$ leaving the AFM ground state undisturbed. 

As the hole doping increases, Fig.~\ref{Shaking} shows that the response to the out-of-phase modulation moves to lower energies. This is caused by two effects: 
First, the energy cost of creating  magnons decreases with doping as we saw above; second, a new low-energy continuum emerges from the rest of the terms in Eq.~\eqref{Shaking} describing the scattering of magnons and holes. Like for $\delta=0$, the 
response is suppressed for small energies and is maximum  close to the energy required to create two ${\bf k}=(\pi,0)$ magnons with maximal energy in the BZ, which decreases with doping as discussed above. The peak also broadens and decreases in height 
with  doping due to damping  of the magnons. 
Figure~\ref{fig:Modulation}  shows that 
there is qualitative agreement between our theory and  the experimental results of Ref.~\cite{kendrick2025pseudogapfermihubbardquantumsimulator}. 
The experimental response is generally broader and at lower energies, which  likely is because it is measured 
 at a non-zero temperature where there is no long range AFM order. However, the energies giving the peak experimental and theoretical response 
 exhibit essentially the same  relative decrease with doping  as shown
in Fig.~\ref{fig:Magnonspectral}(c). The response to the in-phase modulation increases from zero with doping and is generally much weaker and broader
than the out-of-phase response 
with no dominant peak. Again, this is consistent with the experimental results where this difference  was interpreted as signs of the 
pseudogap. The theoretical  response can in fact be matched even better with the experimental data using a simple rescaling of the spin coupling $J$, see Supplemental Material \cite{SM}.
Importantly, the scaling factor of $J$ is the same for the in-phase and the 
out-of-phase modulation and for all dopings. It can therefore be regarded as  an ad-hoc way to include temperature effects. As for the spin correlations discussed above, these results indicate that the observed $T>0$ response is inherited from an AFM ground state. 
The agreement between theory and experiment breaks down for larger doping where the system is expected to enter a cross-over region to a Fermi liquid, Supplemental Material~\cite{SM}. 

\paragraph{Discussion and conclusions.--}
We used a conserving diagrammatic approach based on a self-consistent Born approximation combined with a self-consistent random phase approximation to analyze the $t-J$ model for small hole doping, where the ground state is an AFM. Doping was shown to lead to damped magnetic polarons forming four elliptical hole pockets in the BZ, and the magnon spectrum was likewise shown to be softened and  damped. The AFM spin correlations were demonstrated  to decrease with hole doping, and  we also calculated the response of the system to  lattice modulations finding a qualitative difference between an in-phase and an out-of-phase modulation in the  $x$- and $y$-directions. Our theoretical results were shown to agree with results from two recent quantum simulation experiments, which were both interpreted as being connected to  pseudogap physics. 

From a broader perspective, the interplay between charge  and spin dynamics, and in particular the physical origin of the pseudogap phase is presently highly debated~\cite{Keimer2015,Nyhegn2025,Bonetti2026}. Our results suggest that this question, central for our understanding of high $T_c$ superconductivity, may be fruitfully approached from the small hole doping side where systematic theories can be developed. An interesting but also challenging future research direction is, therefore, to generalize our theoretical framework to the case of a non-zero temperature where there is no long-range AFM order. Another intriguing question concerns the transition between the small doping regime where magnetic polarons are the main charge carriers, to the large doping regime where the system is a Fermi liquid of electrons~\cite{Koepsell2021}, which will require including terms beyond the regime of LSWT and  small hole doping. A conserving calculation of the response to lattice modulations is also an important  but complicated problem, and it would finally be highly interesting to explore possible Cooper pairing instabilities as a function of doping. \\

\begin{acknowledgments}
G.M.B.\ thanks Martin Lebrat and Eugene Demler for useful discussions. M.C.\ and G.M.B.\ acknowledge support from the Novo Nordisk Foundation (grant no.\ NNF23OC0086599). K.K.N. acknowledges support from the Carlsberg Foundation through a Carlsberg Reintegration Fellowship (Grant no. CF24-1214). J.H.N acknowledge support from the Independent Research Fund Denmark (grant no. 5341-00014B).
\end{acknowledgments}

\bibliographystyle{apsrev4-2}
\bibliography{ref_MagnonFiniteDens}

\end{document}


\title{Supplemental Material \\ Hole and spin dynamics in an anti-ferromagnet close to half filling}
\maketitle
\beginsupplement
\tableofcontents

\section{A: Luttinger-Ward functional for SCBA+RPA} 

\begin{figure}[ht!]
    \centering
    \includegraphics[width=0.5\columnwidth]{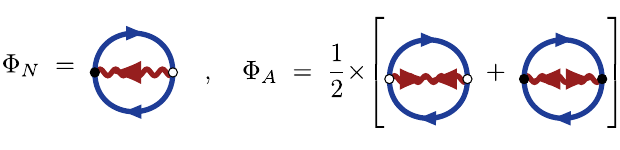}
   \caption{Normal ($\Phi_N$) and anomalous ($\Phi_A$) contributions to the Luttinger-Ward functional, $\Phi = \Phi_N + \Phi_A$, for the SCBA+RPA.}
    \label{fig:LW_functional} 
\end{figure}

\begin{figure}[h!]
    \centering
    \includegraphics[width=0.5\columnwidth]{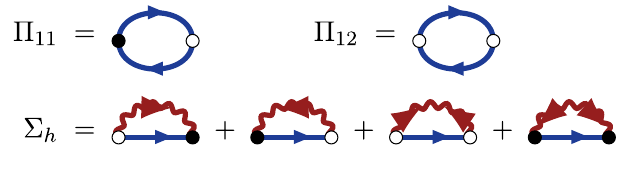}
    \caption{The magnon self-energies (top) in the self-consistent RPA, and the hole self-energy (bottom) in the SCBA. These are derived from the Luttinger-Ward functional in Fig.~\ref{fig:LW_functional}.}
    \label{fig:self_energies}
\end{figure}

Here, we construct the Luttinger-Ward functional that gives rise to the self-consistent Dyson equations in Fig.~1(a) of the main text. We start from the Dyson equations
\begin{align}
G(k) &= G_0(k) + G_0(k) \Sigma_h(k) G(k),  \; \mathbf{D}(k) = \mathbf{D}_0(k) + \mathbf{D}_0(k) \mathbf{\Pi}(k) \mathbf{D}(k).
\end{align}
Here, $k = (\mathbf{k},i\omega_k)$ and we define the propagators as in the main text, with 
\begin{align}
\mathbf{D}(k) &= \begin{bmatrix} D_{11}(k) & D_{12}(k) \\ D_{21}(k) & D_{22}(k)\end{bmatrix}, \; \mathbf{\Pi}(k) = \begin{bmatrix} \Pi_{11}(k) & \Pi_{12}(k) \\ \Pi_{21}(k) & \Pi_{22}(k)\end{bmatrix}.
\end{align}
Using that $D_{12}^0 = D_{21}^0 = 0$, we obtain the scalar Dyson equations for the magnons
\begin{align}
D_{11}(k) &= D_{11}^0(k) \left[1 +  \Pi_{11}(k) D_{11}(k) + \Pi_{12}(k) D_{21}(k)\right], \nonumber \\
D_{12}(k) &= D_{11}^0(k) \left[\Pi_{11}(k) D_{12}(k) + \Pi_{12}(k) D_{22}(k)\right].
\end{align}
Now, we define the Luttinger-Ward functional \cite{Baym1961,Baym1962, Stefanucci_vanLeeuwen_2025,Secchi2020} as the sum of the two terms, $\Phi = \Phi_N + \Phi_A$, in Fig.~\ref{fig:LW_functional}. Explicitly,
\begin{align}
\Phi_N =&\, \frac{1}{\beta^2}\sum_{p,q} g^2(\mathbf{p},\mathbf{q}) G(p) D_{11}(-q) G(p+q),\nonumber \\
\Phi_A =&\, \frac{1}{2\beta^2}\sum_{p,q} g(\mathbf{p},\mathbf{q}) g(\mathbf{p}+\mathbf{q},-\mathbf{q}) G(p)  \left[D_{12}(-q) + D_{21}(+q)\right] G(p+q).
\end{align}
From this, we derive the hole self-energy as the functional derivative with respect to $G(k)$ [intuitively, this corresponds to cutting a hole propagator line in $\Phi$]
\begin{align}
\Sigma_h(k) = -\beta \frac{\delta \Phi}{\delta G(k)} =&\, -\frac{1}{\beta}\sum_q  G(k \!+\! q)  \left[g^2(\mathbf{k} \!+\! \mathbf{q},-\mathbf{q}) D_{11}(q) \!+\! g^2(\mathbf{k},\mathbf{q}) D_{11}(-q)\right] \nonumber \\
&\, - \frac{1}{\beta}\sum_q  g(\mathbf{k},\mathbf{q}) g(\mathbf{k} + \mathbf{q},-\mathbf{q}) G(k \!+\! q)  \left[ D_{12}(-q) + D_{21}(+q)\right].
\end{align}
Similarly, the magnon self-energies are [again, intuitively corresponding to cutting magnon propagator lines in Fig.~\ref{fig:LW_functional}]
\begin{align}
\Pi_{11}(k) =&\, \beta \frac{\delta \Phi}{\delta D_{11}(k)} = \frac{1}{\beta}\sum_p  g^2(\mathbf{p},-
\mathbf{k}) G(p) G(p - k). \nonumber \\
\Pi_{12}(k) =&\, \beta\left(\frac{\delta \Phi}{\delta D_{12}(k)} \!+\! \frac{\delta \Phi}{\delta D_{12}(-k)}\right) = \frac{1}{2\beta} \sum_p \Big(g(\mathbf{p},-\mathbf{k}) g(\mathbf{p} \!-\! \mathbf{k},\mathbf{k}) G(p) G(p \!-\! k) \!+\! g(\mathbf{p},\mathbf{k}) g(\mathbf{p} \!+\! \mathbf{k},-\mathbf{k}) G(p) G(p \!+\! k)\Big) \nonumber \\
=&\, \frac{1}{\beta} \sum_p g(\mathbf{p},\mathbf{k}) g(\mathbf{p} + \mathbf{k},-\mathbf{k}) G(p) G(p + k).
\end{align}
Note that the anomalous magnon self-energy $\Pi_{12}$ follows from a \emph{symmetrized} derivative. This comes about, because of the symmetry $D_{12}(k) = D_{12}(-k)$, whereby one cannot independently vary with respect to  $D_{12}(k)$ and $D_{12}(-k)$ \cite{Stefanucci_vanLeeuwen_2025}. These self-energies correspond to the self-consistent Born approximation (SCBA) for the holes and the self-consistent rotating phase approximation (RPA) for the magnons, see Fig.~\ref{fig:self_energies}. As they are derived from a Luttinger-Ward functional, the applied SCBA+RPA scheme is a conserving approximation \cite{Baym1961,Baym1962, Stefanucci_vanLeeuwen_2025}.

\section{B: Spin Correlations}
\begin{figure}[t!]
    \centering
    \includegraphics[width=0.6\columnwidth]{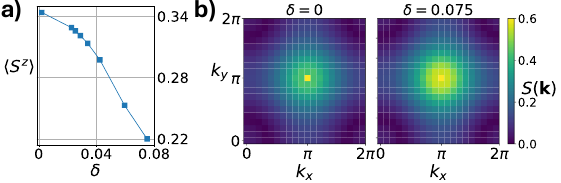}
    \caption{(\textbf{a}) The staggered magnetisation for small dopings. (\textbf{b}) The spin structure factor
    for two different dopings. 
    }
    \label{fig:Mag_App}
\end{figure}

Defining the staggered magnetisation as 
$M_{stag}=\frac{1}{N}\sum_i\eta_i S^z_i$ for $T=0$ with $\eta_i$ being alternatively $1$ and $-1$ for the A and B sublattice respectively, we find
\begin{equation}
\begin{aligned}
    M_{stag} = \frac{1}{2}\left(1-\delta\right)
    -\frac{1}{N}\sum_{\mathbf{k}} \bigg[ v_{\mathbf{k}}^2 + n_\mathbf{k}^{11}\left(u_\mathbf{k}^2 + v_\mathbf{k}^2\right) 
    -2u_\mathbf{k}v_\mathbf{k}n_\mathbf{k}^{12} \bigg]= \frac{1}{2}\left(1-\delta\right) - S_0.
\end{aligned}
\end{equation}
Here, $n_{\mathbf{k}}^{i}=-\int_{-\infty}^0 A_{i}(\textbf{k}, \omega)d\omega/(2\pi)$ for $i\in [ 11, 12]$, and $S_0=\sum_\textbf{k}S_v(\textbf{k})=\sum_\textbf{k}(v_\textbf{k}^2 +n_\textbf{k}^{11}(u_\textbf{k}^2 +v_\textbf{k}^2)-2u_\textbf{k}v_\textbf{k}n_\textbf{k}^{12})$. 
The staggered magnetisation is plotted against doping in Fig.\ \ref{fig:Mag_App}. We clearly see how it decreases with $\delta$
as expected. 
For $\delta=0$, we find $\left<S^z\right>\simeq0.35$ which is larger than the usual LSWT result $\left<S^z\right>\simeq0.303$
obtained in the limit $N\rightarrow\infty$ of an infinite lattice \cite{Auerbach_book}. 
This difference is due to finite size effects in our $20\times20$ lattice.

We now calculate the longitudinal correlation function $4\left< S^z(\mathbf{d})S^z(\mathbf{0})\right>=\frac{4}{N}\sum_i\left< S_{i+\mathbf{ d}}^zS_i^z\right>$. 
Since the Pauli matrices satisfy $\sigma_i^2=\mathbb{I}$, this gives  $1-\delta$ for  $\mathbf{d} =  \mathbf{0}$. For $\mathbf{d} \neq \mathbf{0}$, we apply the Boguliobov transformations and 
afterwards Wick's theorem. After some algebra one finds 
\begin{equation}
\begin{aligned}
    4\left<S^z(\mathbf d)S^z(\mathbf{0})\right>&\simeq4(-1)^\textbf{d}\bigg[\frac{1}{4}\left(1-\delta\right)^2-S_0(1-\delta-S_0)\quad+ \sum_{\textbf{k}\textbf{q}}e^{i(\textbf{k}-\textbf{q})\textbf{d}}\Tilde{S}(\textbf{k})\Tilde{S}(\textbf{q}) +\sum_{\textbf{k}\textbf{q}}e^{i(\textbf{k}-\textbf{q})\textbf{d}} S_u(\textbf{k})S_v(\textbf{q}) 
    \\
    &-\frac{1}{4}|H(\textbf{d})|^2\bigg]
    \label{szz}
\end{aligned}
\end{equation}
with
\begin{align*}
    H(\textbf{d}) &= \frac{1}{N}\sum_\textbf{k}\left<h^\dagger_\textbf{k}h_\textbf{k}\right>e^{i\mathbf{k}\cdot\mathbf{d}}, 
    \quad \quad
    \Tilde{S}(\textbf{k}) = \frac{1}{N}\left[-u_\textbf{k}v_\textbf{k}(2n_\textbf{k}^{11}+1) + n_\textbf{k}^{12}(u_\textbf{k}^2 + v_\textbf{k}^2) \right]
    \\
    S_u(\textbf{k}) &= \frac{1}{N}\left[u_\textbf{k}^2 + n_\textbf{k}^{11}(u_\textbf{k}^2+v_\textbf{k}^2)-2u_\textbf{k}v_\textbf{k}n_\textbf{k}^{12}\right],\quad \text{and} \quad
    S_v(\textbf{k}) = \frac{1}{N}\left[v_\textbf{k}^2 + n_\textbf{k}^{11}(u_\textbf{k}^2+v_\textbf{k}^2)-2u_\textbf{k}v_\textbf{k}n_\textbf{k}^{12}\right].
\end{align*}
Subtracting the mean we find 
\begin{align*}
    C^z(\mathbf{d}) &= 4\left[\left<S^z(\mathbf{d})S^z(\mathbf{0})\right>-\left<S^z(\mathbf{d})\right>\left<S^z(\mathbf{0})\right>\right]
    \\
    &=4\left[\sum_{\textbf{k}\textbf{q}}e^{i(\textbf{k}-\textbf{q})\cdot\textbf{d}}\Tilde{S}(\textbf{k})\Tilde{S}(\textbf{q}) +\sum_{\textbf{k}\textbf{q}}e^{i(\textbf{k}-\textbf{q})\cdot\textbf{d}} S_u(\textbf{k})S_v(\textbf{q}) 
    - \frac{1}{4}|H(\textbf{d})|^2\right]
\end{align*}
By applying inversion symmetry one can see that $C^z(\textbf{d})$ indeed is real. Taking the Fourier transform of $\left<S^z(\mathbf{d})S(\mathbf{0})\right>$ we determine the spin structure factor $S(\textbf{k})=\frac{1}{N}\sum_\textbf{d} \left<S^z(\mathbf{d})S(\mathbf{0})\right>e^{i\textbf{k}\cdot\textbf{d}}$. Note that this approach using Wick's theorem corresponds to LSWT, which ignores interactions between the magnons.

We plot in Fig.\ \ref{fig:Mag_App}(b) the spin structure factor  at half-filling and at doping $\delta=0.075$. We clearly see a peak at the AFM 
ordering vector $(\pi, \pi)$ for both dopings. For $\delta=0.075$, the peak is however broader and significantly lower (reduced by a factor $\sim 0.56$) in agreement with what is observed experimentally~\cite{Chalopin2026}.

We also calculate the transverse correlation function $C^x(\mathbf{d})=C^y(\mathbf{d})=\frac{4}{N}\sum_i\left<S^x_{i+\mathbf{d}}S^x_i\right>$, which is different from the longitudinal 
correlation function in the broken symmetry phase. In addition, we of course have  $\langle S^x_i\rangle=\langle S^y_i\rangle=0$. 
As above, we use the Boguliubov transformation together with Wick's theorem ignoring interactions between the magnons, which corresponds 
to LSWT.  Including terms up to and including quartic order gives 
\begin{align*}
     C^x(\mathbf{d}) &= \frac{2(1-2\delta)}{N}\sum_\textbf{k}\cos(\textbf{k}\cdot\textbf{d})\Psi_\textbf{k}\left[n_\textbf{k}^{11}+n_\textbf{k}^{12}+1/2\right] 
     +\frac{4}{N}\sum_{\textbf{k}'}\cos(\textbf{k}'\cdot \textbf{r})S_v(\textbf{k}')
     \\
     &-\frac{4}{N}\sum_{\textbf{k}'\textbf{k}}\cos(\textbf{k}'\cdot\textbf{r})\bigg[2\Tilde{S}(\textbf{k}')S_v(\textbf{k}) +S_v(\textbf{k}')\Tilde{S}(\textbf{k})
     +S_v(\textbf{k}')S_v(\textbf{k})+\Tilde{S}(\textbf{k}')\Tilde{S}(\textbf{k})+S_v(\textbf{k}')S_u(\textbf{k})\bigg]
\end{align*}
with $\Psi_\textbf{k}= [(1 - \gamma_\textbf{k})/(1 + \gamma_\textbf{k})]^{1/2}$. To compare with experiments where 
the magnetic order points in random directions from shot to shot, we  define the average correlation 
function $\mathcal{C}_2(\mathbf{d})=\frac{1}{3}C^z(\mathbf{d}) + \frac{2}{3}C^x(\mathbf{d})$. 


\section{C: Lattice Modulation}
In this section, we consider the response of the system  to an in-phase ($T_+$) and out-of-phase ($T_-$) lattice modulation
in the $x$- and $y$-directions 
as measured recently~\cite{kendrick2025pseudogapfermihubbardquantumsimulator}. This experiment  measures the heating rate caused by the lattice modulation in
a sample of ultra-cold ($T/t<0.15$) $^6$Li atoms realising a nearest neighbour Hubbard model with  $U=7t$ and dopings $0\le \delta\le 0.30$. 
The  perturbation caused by the lattice modulation is 
\begin{equation}
\begin{aligned}
    H'(\tau) &= V\cos(\omega\tau)T_{\pm}
    \quad\quad \text{with}\quad\quad  T_{\pm} = \sum_{i\sigma}c_{i\sigma}^\dagger c_{i+x\sigma}\pm c_{i\sigma}^\dagger c_{i+y\sigma} +\text{h.c.}
    \label{lattice modulation}
\end{aligned}
\end{equation}
This gives rise to the heating rate~\cite{kendrick2025pseudogapfermihubbardquantumsimulator}
\begin{align*}
    \left|\overline{\frac{dQ}{d\tau}}\right| = \frac{1}{2}\omega|V|^2 \text{Im}[\chi_-(\omega)]
     \quad\quad \text{with}\quad\quad
    \chi_{\pm}(\omega) = -i\int_{-\infty}^{\infty}\Theta(t') \left<\left[T_\pm(t'), T_\pm(0)\right]\right>e^{-i\omega t'}dt'.
\end{align*}

Care has to be taken when calculating $\chi_{\pm}$ from the $t-J$ model, since the mapping from the Hubbard 
to the $t-J$ model  involves a unitary Schrieffer-Wolff transformation $e^S$, to 
systematically include doubly occupied states to lowest order~\cite{schrieffer1966anderson}. 
Specifically, the $t-J$ model is obtained from the Hubbard model $\hat H=\hat H_t+\hat H_U$ with 
$H_t = -t\sum_{<i,j>}c_{i\sigma}^\dagger c_{j\sigma} +\text{h.c.}$ and $H_U=U\sum_i n_{i,\uparrow}n_{i, \downarrow}$ 
by the transformation $P_L e^SHe^{-S}P_L$. Here $P_L$ projects onto states with maximally one particle per lattice site and 
$S$ is  chosen such that $P_Le^SHe^{-S}P_H=0$ to leading order in $t$. Defining $P_L = \Pi_i(1-n_{i,\uparrow}n_{i, \downarrow})$ and $P_H=1-P_L$, we may write $S=-P_HH_U^{-1}H_tP_L - P_LH_tH_U^{-1}P_H$ ~\cite{Auerbach_book}. 
The point is that this transformation also changes the operator $T_{\pm}$ to 
$T_\pm^\text{eff}=P_Le^S T_\pm e^{-S}P_L$. To first order in $t/U$, we find 
\begin{align*}
    T_\pm^\text{eff} \approx P_L \left(T_\pm + [S, T_\pm]\right)P_L = T_\pm^0 + T_\pm^1.
\end{align*}
Rewriting this using the  slave fermion representation and ignoring quartic terms then gives  Eq.~(5) in the main text. 
 
\begin{figure}[t!]
    \centering
    \includegraphics[width=\columnwidth]{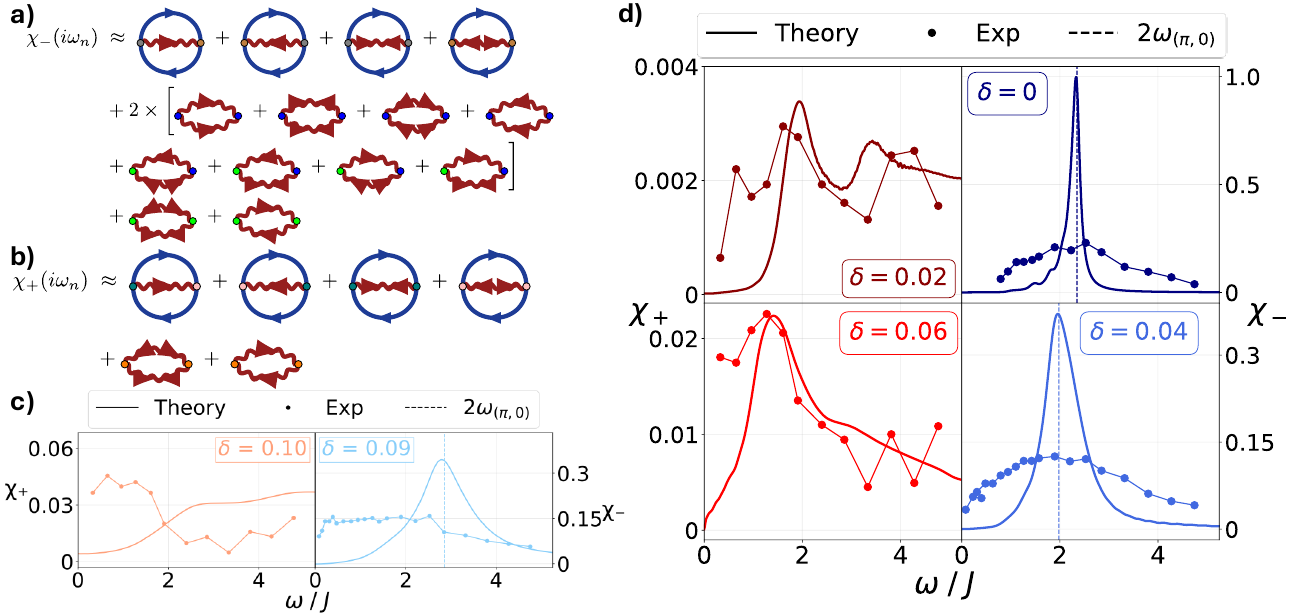}
    \caption{(\textbf{a}) Diagrams corresponding to our calculation of the  out-of-phase  response function $\chi^\text{mat}_-$. 
    Grey vertices correspond to the factor $g^-(\mathbf{q}, \mathbf{k})$, where $\mathbf{q}$ is the incoming hole momentum and $-\mathbf{k}$ is the outgoing magnon momentum. 
    Brown vertices correspond to the factor $g^-(\mathbf{k}+\mathbf{q}, -\mathbf{k})$, where $\mathbf{k}$ and $\textbf{q}$ is the incoming magnon and hole momentum respectively. 
    Blue vertices correspond to the factor  ${2J\tilde{\gamma}_\textbf{k}}/({t\sqrt{1-\gamma_\textbf{k}^2}})$ and green vertices to
    $-{4\gamma_\textbf{k}J\tilde{\gamma}_\textbf{k}}/({t\sqrt{1-\gamma_\textbf{k}^2}})$. For $T=0$ and $\delta=0$, only the  first  diagram in the second line is non-zero, and it gives rise to the two magnon peak. 
    (\textbf{b}) Diagrams corresponding to our calculation of the in-phase response function $\chi^\text{mat}_+$.
    Teal vertices correspond to the factor $g^+(\mathbf{q}, \mathbf{k})$, where $\mathbf{q}$ is the incoming hole momentum and $-\mathbf{k}$ is the outgoing magnon momentum. 
    Pink vertices  correspond to the factor $g^+(\mathbf{k}+\mathbf{q}, -\mathbf{k})$ where $\mathbf{k}$ and $\textbf{q}$ is the incoming magnon and hole momentum respectively. 
    Orange vertices  correspond to the factor ${2\omega_\textbf{k}}/{t}$. All diagrams vanish for $T=0$ and $\delta=0$. 
    (\textbf{c}) Response to an in-phase (left) and out-of-phase (right) lattice modulation for dopings  $\delta=0.1$ and $\delta=0.09$. 
    The experimental points are taken from Ref. \cite{kendrick2025pseudogapfermihubbardquantumsimulator}. The calculation presented here is, unlike the other calculations presented in the article, made for a $16\times16$ grid with $\eta=0.08$.
    (\textbf{d}) A calculation of the in-phase and out-of-phase lattice modulation responses for the same dopings as in Fig. 4 of the main text, but where we have rescaled 
    $J\rightarrow (2.35/3.3)J$.  
    }
    \label{fig:LM}
\end{figure}

Having derived $T_\pm^\text{eff}$, 
we then consider the Matsubara response function $\chi_\pm^\text{mat}(\tau)=-\left<T_\tau[T^{eff}_\pm(\tau),T^{eff}_\pm(0)]\right>$ with 
$\tau=it'\in [-1/T;1/T[$  imaginary time~\cite{BruusFlensberg}. The retarded correlation function  can then be obtained from the 
Fourier transform
\begin{align}
    \chi^\text{mat}_\pm(i\omega_n)= \int_0^\beta e^{i\omega_n\tau}\left<T_\pm^{eff}(\tau)T_\pm^{eff}(0)\right>d\tau\simeq\chi_\pm^{00}+\chi_\pm^{01}+\chi_\pm^{10}+\chi_\pm^{11},
    \label{Response}
\end{align}
by the analytic continuation $\chi_{\pm}(\omega)=\chi^\text{mat}_\pm(i\omega_n\rightarrow \omega+i0_+)$. Here  $\omega_n=2nT$ is a bosonic Matsubara frequency.
In the last equality of Eq.~\eqref{Response},  
 we have used Wick's theorem corresponding to LSWT.
 We furthermore assume  a small hole concentration  $\delta<J/t$
 so that the term $\chi_\pm^{00}$ scaling as first order in $\delta$ and the term $\chi_\pm^{11}$ scaling as 
  second order in $t/U$ are included whereas the terms  $\chi_\pm^{01}$ and $\chi_\pm^{10}$ scaling as 
 $\delta^2$ or $\delta t/U$ are ignored. 
Equation \eqref{Response}  is illustrated diagrammatically in Fig.~\ref{fig:LM}. While this approximation includes the effect of  magnon-hole interactions by using fully dressed 
Green's functions, it would be very interesting but also challenging to go beyond  the small  doping assumption and  LSWT  to include magnon-magnon interactions 
in a conserving calculation of the response.

In Fig. \ref{fig:LM}(c), we present the response to an in- and an out-of-phase  lattice modulation for  dopings $\delta=0.1$ and $\delta=0.09$ respectively. 
There is less agreement between theory and experiment than for the dopings in the main manuscript. We speculate that this is due to the system 
being in the cross-over region between the pseudogap and the Fermi liquid phase for these higher hole dopings, where we do not expect our small doping theory to apply~\cite{kendrick2025pseudogapfermihubbardquantumsimulator}. 

In Fig. \ref{fig:LM}(d), we show results for the same dopings as in the main manuscript but where we have now tuned the spin coupling coefficient $J$ 
in the calculations to fit the theoretically predicted maxima with the experimental data. Importantly,  we use the same rescaling of $J$ for the in-phase and 
out-of-phase modulations and for all dopings. We see that there are now even better agreement between theory and 
experiment, and this rescaling procedure can therefore be regarded as an ad hoc way to incorporate finite temperature effects. 

\section{D: Numerical analysis}
Achieving self-consistent solutions for the Green's functions in this article is computationally expensive and the convergence is slow. We have used differentiated frequency resolutions with finest resolution for $\omega \in [-t, 3t ]$. For most of our calculations, the grid has been truncated outside $-6t$ and $8t$. To ensure that 
peaks are fully captured, we have used the frequency steps $\Delta \omega < \eta/2$. 
In all calculations we have taken $\eta=0.04t$ except in Fig.~\ref{fig:LM}(c) where have used $\eta=0.08t$.
To achieve convergence we have updated the self-energies partially so  that the self-energy $\Sigma^n$ at iteration $n$ is given by 
$\Sigma^n = (1-\alpha)\Sigma^{n-1} + \alpha f(\Sigma)$ where $f(\Sigma)$ is the result from calculating the self-energy expressions and 
 $\alpha\simeq 0.10$. Similar computational tricks have been applied in Refs.~\cite{Satyaki2011, sherman1997tJ}. 
 For all integrals we have used the trapezoidal integration method and exploited symmetries such as $D_{12}(\textbf{k}, iq_n)=D_{21}(\textbf{k}, iq_n), (D_{11}^A)^\dagger = D_{11}^R, (G_h^A)^\dagger = G_h^R$, and $(D_{22}^A)^\dagger = D_{22}^R$, which are preserved through the iteration process~\cite{sherman1997tJ}. We have also used that all Greens functions are invariant under $C_4$ transformations and under the transformation $\textbf{k}\rightarrow \textbf{k}+(\pi, \pi)$, with the exception of $D_{12}(\textbf{k}, i\omega_n)$ where $D_{12}(\textbf{k}+(\pi, \pi), i\omega_n)=-D_{12}(\textbf{k}, i\omega_n)$. This allows us to restrict the $20\times20$ lattice points to 36 unique momentum states. For pairs of momentum states related both by a $C_4$ transformation and $\textbf{k}\rightarrow \textbf{k}+(\pi, \pi)$ up to an integer multiple of reciprocal lattice vectors, such as $(\pi, 0)$ and $(0, \pi)$ or $(\pi/2, \pi/2)$ and $(3\pi/2, 3\pi/2)$, we find $D_{12}(\textbf{k}, i\omega_n)=0$.
\newpage

\bibliographystyle{apsrev4-2}
\bibliography{ref_MagnonFiniteDens}